\documentclass[a4paper]{jpconf}
\usepackage{amsfonts}
\usepackage{graphicx}
\begin{document}
\title{Factorization of supersymmetric Hamiltonians in curvilinear coordinates}

\author{M A Gonz\'alez Le\'on$^1$, J Mateos Guilarte$^2$ and M de la Torre Mayado$^2$}

\address{$^1$ Departamento de Matem\'atica Aplicada and {\sl IUFFyM}. Universidad de Salamanca, Spain}

\address{$^2$ Departamento de F\'{\i}sica Fundamental and {\sl IUFFyM}. Universidad de Salamanca, Spain}

\ead{magleon@usal.es, guilarte@usal.es, marina@usal.es}

\begin{abstract}
Planar supersymmetric quantum mechanical systems with separable spectral problem in curvilinear coordinates are analyzed in full generality. We explicitly construct the
supersymmetric extension of the Euler/Pauli Hamiltonian describing the motion of a
light particle in the field of two heavy fixed Coulombian centers. We shall also show
how the SUSY Kepler/Coulomb problem arises in two different limits of this problem: either, the two
centers collapse in one center - a problem separable in polar coordinates -, or, one of the two centers flies to infinity
- to meet the Coulomb problem separable in parabolic coordinates.
\end{abstract}

\section{Introduction}

The classification of quantum Hamiltonians for which the Scr$\ddot{\rm o}$dinger equation is
separable in curvilinear coordinates is well known. Eisenhart in Reference \cite{Eisenhart} completely achieved this task in three dimensional systems. More recently Andrianov and Ioffe worked this problem in supersymmetric quantum mechanical systems both in two and three dimensions \cite{Ioffe}. Our aim in this paper is to describe how SUSY quantum mechanical systems separable in polar, elliptic, and parabolic coordinates can be constructed in a rather geometrical way. We shall follow previous work of our group on the interplay of integrability and supersymmetry in some planar supersymmetric systems, both classical and quantal, see \cite{AoP, JPA, Sigma, CM}. In particular, we shall address a SUSY model built from the Euler two-center problem which is separable in elliptic coordinates. As a bonus, we shall show that the planar SUSY Kepler/Coulomb problem built by Ioffe and collaborators in \cite{Ioffe1} and independently by Wipf el al in \cite{Wipf} arises in two different limits corresponding respectively to the separability of the Kepler/Coulomb problem in polar or parabolic coordinates.

\section{${\cal N}=2$ Supersymmetric Quantum Mechanics}

\subsection{${\cal N}=2$ Classical Supersymmetric mechanics}

We start by describing ${\cal N}=2$ SUSY classical systems such that the configuration space is a two-dimensional Riemannian manifold $({\mathbb M}^2,g_{\mu\nu})$. Let $q^\mu$, $\mu=1,2$, be a system of local coordinates in ${\mathbb M}^2$ and $\vartheta_\alpha^\mu$, $\alpha,\mu\, =\, 1,2$, a set of Grassman variables. A super-point in the configuration super-space ${\cal C}\cong {\mathbb M}^{2|4}$ is given by: $(q^\mu,\vartheta^\mu_\alpha)\equiv
(q^1,q^2,\vartheta_1^1,\vartheta_1^2,\vartheta_2^1,\vartheta_2^2)$, whereas the following commutation/anti-commutation rules hold:
\[
q^\mu \vartheta^\nu_\alpha -\vartheta^\nu_\alpha q^\mu = 0\quad , \quad  \vartheta_\alpha^\mu
\vartheta_\beta^\nu + \vartheta_\alpha^\nu
\vartheta_\beta^\mu = 0 \quad .
\]
The super-symmetric action may be expressed in the form (Einstein summation convention will be used along the paper):
\[ S = \int\, d t \, \left\{
\frac{1}{2} g_{\mu\nu}\ \dot{q}^\mu\  \dot{q}^\nu  +  \frac{i}{2} g_{\mu\nu}\
\vartheta_{\alpha}^\mu D_t \vartheta_{\alpha}^\nu + \frac{1}{4} R_{\mu\nu\rho\eta}
\vartheta_1^\mu \vartheta_2^\nu \vartheta_1^\rho \vartheta_2^\eta  -\frac{1}{2} g^{\mu\nu}\frac{\partial W}{\partial q^\mu} \frac{\partial W}{\partial q^\nu} + i \frac{\partial^2 W}{\partial q^\mu
\partial q^\nu} \vartheta_1^\mu \vartheta_2^\nu \right\} \quad .
\]
Here: $W(q^1,q^2): {\mathbb M}^2 \rightarrow {\mathbb R}$, is the superpotential that characterizes the interactions compatible with supersymmetry in the system. The covariant derivative for the Grassman variables is
\[
D_t \vartheta_\alpha^\mu=\dot{\vartheta}_\alpha^\mu+\Gamma^\mu_{\nu \rho}
\dot{q}^\rho \vartheta_\alpha^\nu
\]
and $\Gamma^\mu_{\nu \rho}$ and $R_{\mu\nu\rho\eta}$ are the standard Christoffel symbols and curvature tensor associated to the metric $g$. Choosing a zweibein $e^\mu_a$ in order to decompose the metric tensor $g^{\mu\nu}$, i.e., $g^{\mu\nu}=e^\mu_a e^\nu_b  \delta^{ab}$, curved and flat Grassman variables can be related in a simple way: $\vartheta^\mu_\alpha=e^\mu_a\, \theta^a_\alpha$.

The action functional is invariant with respect to the SUSY transformations
\[ \delta_1 q^\mu= \varepsilon \vartheta_1^\mu \  , \  \delta_1
\vartheta_1^\mu= i \varepsilon \dot{q}^\mu \  , \  \delta_1
\vartheta_2^\mu=i \varepsilon \left(g^{\mu\nu} \frac{\partial W}{\partial
q^\nu}+i\Gamma^\mu_{\nu\rho}\vartheta_1^\nu\vartheta_2^\rho\right)
\]
\[
\delta_2 q^\mu= \varepsilon \vartheta_2^\mu \ , \ \delta_2
\vartheta_1^\mu=- i \varepsilon \left(g^{\mu\nu} \frac{\partial
W}{\partial q^\nu} + i\Gamma^\mu_{\nu\rho}\vartheta_1^\nu\vartheta_2^\rho\right) \
, \ \delta_2 \vartheta_2^\mu= i \varepsilon \dot{q}^\mu
\]
where $\varepsilon$ is a Grassman parameter, and the N\oe ther's theorem gives the conserved super-charges:
\begin{equation}
Q_1=g_{\mu\nu} \dot{q}^\mu \vartheta_1^\nu +
\frac{\partial W}{\partial q^\mu} \vartheta_2^\mu\,\, ,\quad
Q_2=g_{\mu\nu} \dot{q}^\mu \vartheta_2^\nu - \frac{\partial W}{\partial q^\mu}
\vartheta_1^\mu
\end{equation}
As a pre-quantization step we pass to the Hamiltonian formalism. A point in the reduced phase super-space is given by: $
(q^\mu , p_\nu, \vartheta^\mu_1, \vartheta^\mu_2)$. In this space a super-Poisson structure is defined in the form:
\[
\{ F, G \}_P = \left( {\displaystyle \frac{\partial F}{\partial
p_\mu} \frac{\partial G }{\partial q^\mu} - \frac{ \partial F}{\partial q^\mu}
\frac{\partial G}{\partial p_\nu}} \right) + { i g^{\mu\nu} \delta^{\alpha\beta}F \frac{
\stackrel{\leftarrow}{\partial}}{\partial \vartheta_{\alpha}^\mu }\frac{
\stackrel{\rightarrow}{\partial}}{\partial \vartheta_{\beta}^\nu } G}  \]
such that $
\{ p_\mu, q^\nu \}_P = \delta_{\mu}^\nu$, $\{ p_\mu, p_\nu\}_P
 = \{ q^\mu, q^\nu \}_P = 0 $, $\{ \vartheta_{\alpha}^\mu, \vartheta_{\beta}^\nu \}_P = i g^{\mu\nu}
\delta_{\alpha\beta}$. It is easy to verify that the supercharges $Q_1$ and $Q_2$ satisfy the classical SUSY algebra: $
\{ Q_1,Q_1 \}_P =  \{ Q_2,Q_2 \}_P =\, 2 i H \, , \,
\{Q_\alpha , H\}_P = \{ Q_1,Q_2 \}_P = 0$, where $H$ denotes the classical SUSY Hamiltonian function of the system. Finally, complex Grassman variables can be defined as follows: $\vartheta_\pm^\mu\, =\, \frac{\pm 1}{\sqrt{2}} (\vartheta_2^\mu\mp i \,\vartheta_1^\mu)$, and the corresponding complex supercharges are written as:
\[
Q_+ = \frac{i}{\sqrt{2}} (Q_1-i Q_2)\quad,\quad  Q_- =  \frac{i}{\sqrt{2}} (Q_1 + i
Q_2)
\]

\subsection{Quantization process}

Canonical quantization in the coordinate representation is performed by the promotions: $( q^\mu, p_\nu) \rightarrow( {\hat q}^\mu,  {\hat
 p}_\nu )$, and $( \vartheta_+^\mu, \vartheta_-^\mu) \rightarrow
 ( {\hat \psi}^{\mu\dag}, {\hat \psi}^\mu)$. ${\hat q}^\mu$ and ${\hat p}^\nu$ become the operators:
\[
{\hat q}^\mu = q^\mu \quad,\qquad  {\hat p}_\nu = {\displaystyle - i \hbar \frac{\partial }{\partial q^\nu}}
\]
whereas the Fermi operators $\hat{\psi}^\mu$ and $\hat{\psi}^{\mu\dag}$ are defined in terms of the zweibein and the \lq\lq planar" Fermi operators $\hat{\psi}^a$ and $\hat{\psi}^{a\dag}$: $
\hat{\psi}^\mu\, =\, e_a^\mu \, \hat{\psi}^a$, $\hat{\psi}^{\mu\dag}\, =\, e_a^\mu \, \hat{\psi}^{a\dag}$.
The quantum operators ${\hat q}^\mu$ and ${\hat p}_\nu$ satisfy the commutation relations:
\[
[\hat{q}^\mu, \hat{p}_\nu ] = i \hbar \delta_\nu^{\mu} \ , \ [ \hat{p}_\mu,
\hat{p}_\nu ] = 0 = [ \hat{q}^\mu, \hat{q}^\nu ]  \ , \ \mu,\nu\, =\, 1,2
\]
and $\hat{\psi}^\mu$ and $\hat{\psi}^{\mu\dag}$ verify the anti-commutation relations of the form:
\[
\{\hat{\psi}^\mu,\hat{\psi}^\nu\}=0=\{\hat{\psi}^{\mu\dag},
\hat{\psi}^{\nu\dag} \}\, , \,
\{\hat{\psi}^\mu,\hat{\psi}^{\nu\dag} \}=\frac{1}{m}g^{\mu\nu}
\]
We will choose a representation for the Fermi operators in terms of the
generators of the Clifford algebra of ${\mathbb R}^4$:
\[
\hat{\psi}^1=\frac{1}{2\sqrt{m}}\left(\gamma^1+i\gamma^3\right)\,  ,
\, \hat{\psi}^2=\frac{1}{2\sqrt{m}}\left(\gamma^2+i\gamma^4\right) \, ,\, \hat{\psi}^{1\dag}=\frac{1}{2\sqrt{m}}\left(\gamma^1-i\gamma^3\right)\,  ,
 \,  \hat{\psi}^{2\dag}=\frac{1}{2\sqrt{m}}\left(\gamma^2-i\gamma^4\right)
\]
where $\gamma^\mu$, $\mu=1,\dots , 4$, are the $4\times 4$  Hermitian
Euclidean gamma matrices:
\[
\gamma^1 = \left( \begin{array}{cc} \sigma^1 & 0 \\ 0 & \sigma^1
\end{array} \right)\
, \  \gamma^2 = \left( \begin{array}{cc} 0 & \sigma^3  \\ \sigma^3 &
0
\end{array} \right) \  , \
\gamma^3 = \left( \begin{array}{cc} \sigma^2 & 0 \\
0 & \sigma^2
\end{array} \right)\  ,\  \gamma^4 = \left( \begin{array}{cc} 0 & \hspace{-0.2cm}-i \sigma^3  \\
i \sigma^3 & 0 \end{array} \right)
\]
$\sigma^i$, $i=1,2,3$, being the Pauli matrices. Thus $\hat{\psi}^\mu$ and $\hat{\psi}^{\mu\dag}$ are explicitly
written as:
\begin{eqnarray*}
&& \hat{\psi}^1 =\frac{1}{\sqrt{m}}\left(
\begin{array}{cccc}
 0 & 1 & 0 & 0  \\
 0 & 0 & 0 & 0  \\
 0 & 0 & 0 & 1  \\
 0 & 0 & 0 & 0
 \end{array}
\right) \,  , \,   \hat{\psi}^2 =\frac{1}{\sqrt{m}}\left(
\begin{array}{cccc}
 0 & 0 & 1 & 0 \\
 0 & 0 & 0 &\hspace{-0.2cm}-1 \\
 0 & 0 & 0 & 0 \\
 0 & 0 & 0 & 0  \\
 \end{array}
\right)\\ &&
\hat{\psi}^{1\dag}
=\frac{1}{\sqrt{m}}\left(
\begin{array}{cccc}
 0 & 0 & 0 & 0  \\
 1 & 0 & 0 & 0  \\
 0 & 0 & 0 & 0  \\
 0 & 0 & 1 & 0
 \end{array}
\right) \,  , \,   \hat{\psi}^{2\dag} =\frac{1}{\sqrt{m}}\left(
\begin{array}{cccc}
 0 & 0 & 0 & 0 \\
 0 & 0 & 0 & 0 \\
 1 & 0 & 0 & 0 \\
 0 & \hspace{-0.2cm}-1 & 0 & 0  \\
 \end{array}
\right) \qquad ,
\end{eqnarray*}
and the quantum supercharges become $4\times4$-matrices of
differential operators:
\begin{equation}
\hat{Q}=\frac{i}{\sqrt{m}}\left(
\begin{array}{cccc}
 0 & D_1-\hbar \omega_1 & D_2 + \hbar \omega_2 & 0 \\
 0 & 0 & 0 & \hspace{-0.2cm} -D_2  \\
 0 & 0 & 0 & D_1 \\
 0 & 0 & 0 & 0 \\
 \end{array}
\right)
\,  ,\
\hat{Q}^\dagger=\frac{i}{\sqrt{m}}\left(
\begin{array}{cccc}
 0 & 0 & 0 & 0  \\
\bar{D}_1 & 0 & 0 & 0 \\
 \bar{D}_2 & 0 & 0 & 0  \\
 0 & \hspace{-0.2cm} -\bar{D}_2-\hbar \omega_2 & \bar{D}_1 + \hbar \omega_1 & 0  \\
 \end{array}
\right)
\end{equation}
Here:
\[
D_\alpha = e_\alpha^\mu \left( \hbar \frac{\partial}{\partial q^\mu} + \frac{\partial
W}{\partial q^\mu} \right)  \quad , \quad  \bar{D}_\alpha = e_\alpha^\mu \left( \hbar \frac{\partial}{\partial q^\mu} - \frac{\partial
W}{\partial q^\mu} \right)\quad ,
\]
and $\omega_1$ and $\omega_2$ denote the {\it \lq\lq spin connection contribution\rq\rq} to the supercharges. For orthogonal metrics, they read:
\[
\omega_1(q^1,q^2) = \frac{1}{2} (g_{11} \Gamma_{22}^1- g_{22}
\Gamma_{12}^2) g^{22} e_1^1\  ,\quad
\omega_2(q^1, q^2)= \frac{1}{2} (g_{11} \Gamma_{21}^1- g_{22}
\Gamma_{11}^2) g^{11} e_2^2
\]
The quantum super-Hamiltonian: $\hat{H}=\frac{1}{2}\{\hat{Q},\hat{Q}^\dagger\}=\frac{1}{2}\left(\hat{Q}\hat{Q}^\dagger+\hat{Q}^\dagger\hat{Q}\right)$, is the $4\times4$ differential operator:
\begin{equation}
\hat{H}=\left(\begin{array}{cccc} \hat{H}_0 & 0 & 0 & 0\\ 0 &
\hat{H}_1^{(11)} & \hat{H}_1^{(12)} & 0 \\ 0 & \hat{H}_1^{(21)} &
\hat{H}_1^{(22)} & 0 \\ 0 & 0 & 0 & \hat{H}_2
\end{array}\right)\label{hamiltonian}
\end{equation}
with a block-diagonal structure inherited from the eigen-spaces of the Fermi number operator:
\[
\hat{N}=\hat{\psi}^\dagger_1\hat{\psi}_1+\hat{\psi}^\dagger_2\hat{\psi}_2=\frac{1}{m}\left(
\begin{array}{cccc}
 0 & 0 & 0 & 0 \\
 0 & 1 & 0 & 0  \\
 0 & 0 & 1 & 0 \\
 0 & 0 & 0 & 2 \\
 \end{array} \right) \qquad .
\]
The vacuum state $|0\rangle$ of the SUSY system is defined by the expressions: $\hat{\psi}^\mu |0\rangle =0\, ,\  \mu=1,2$, and thus the Fock space is built from this state as follows: The creation operators acting on $|0\rangle$ bring the system into one-particle (fermionic) states: $\hat{\psi}^{\mu \dagger}|0\rangle =|1_\mu\rangle$. The two-particle (bosonic) state is: $\hat{\psi}^{2\dagger}|1_1\rangle = |1_1 1_2\rangle
=-\hat{\psi}^{1\dagger}|1_2\rangle=- |1_2 1_1\rangle$. A general state in this space, ${\mathcal F}={\mathcal F}_0\oplus{\mathcal
F}_1\oplus{\mathcal F}_2 $, can be written as a linear combination of the form:
\[
|f\rangle= f_0|0\rangle + \sum_{k=1}^2 \, f_{1k} |1_k\rangle + f_2 |1_11_2 \rangle \  , \quad f_0, f_{1k} , f_2
\in{\mathbb C} \quad .
\]
The supersymmetric space of states is the direct product of ${\mathcal F}$ with  the Hilbert space $L^2({\mathbb R}^2)$:
\[
{\mathcal S}{\mathcal
H}={\mathcal H}\otimes{\mathcal F}=L^2({\mathbb R}^2)\otimes{\mathbb C}^4=
{\mathcal S}{\mathcal H}_0\oplus {\mathcal S}{\mathcal H}_1\oplus{\mathcal S}{\mathcal H}_2
\]
\[ |\Psi(q^1,q^2)\rangle=\, f_0(q^1,q^2)|0\rangle +\sum_{k=1}^2 f_{1k}(q^1,q^2)|1_k\rangle +
 f_{2}(q^1,q^2)|1_1 1_2\rangle
\]
Thus Fermi operators can be viewed as the generators of the
Clifford algebra of ${\mathbb R}^4$ acting on the space of
four-component Euclidean spinors generated by
\[ \left\{
 |0\rangle \equiv \left(\begin{array}{c} 1 \\ 0 \\ 0 \\ 0\end{array}\right)\,  , \,
 |1_1\rangle \equiv \left(\begin{array}{c} 0 \\ 1 \\ 0 \\ 0\end{array}\right)\,  , \,
 |1_2\rangle \equiv  \left(\begin{array}{c} 0 \\ 0 \\ 1 \\ 0 \end{array}\right)\,   , \,
 |1_11_2\rangle \equiv \left(\begin{array}{c} 0 \\ 0 \\ 0 \\ 1 \end{array}\right) \right\}
\]
The supercharges move states between the different Fermionic sectors of the space of states:
\[
{\cal S}{\cal H}_{0}\begin{array}{c} \hat{Q}^\dagger \\ \rightleftharpoons \\
\hat{Q}\end{array} {\cal S}{\cal H}_{1} \begin{array}{c} \hat{Q}^\dagger
\\ \rightleftharpoons \\ \hat{Q} \end{array} {\cal S}{\cal H}_{2}
\]

\section{Separable SUSY Quantum Mechanical systems}

We shall consider ${\cal N}=2$ supersymmetric systems with Liouville models as the bosonic sector. The key property enjoyed by this class of systems is that they are separable using two-dimensional elliptic, polar, parabolic or cartesian systems of coordinates. Thus the target space for these systems is the Riemannian manifold ${\mathbb M}^2$ equipped
with the metric induced from the change of coordinates to two-dimensional elliptic, polar and
parabolic coordinates, and the Euclidean metric for cartesian ones. All these coordinates are orthogonal, i.e.: $g_{\mu\nu}=0 \, ,\, \forall \mu\neq \nu$, and thus there exists a \lq\lq natural" choice of the zweibein:
\[
e^1_1=\sqrt{g^{11}} \ , \
e^1_2=0 \ , \
e^2_1=0 \ , \   e^2_2=\sqrt{g^{22}}\quad .
\]
The super-Hamiltonian (\ref{hamiltonian}) is written in these cases as follows: In the bosonic sectors the Hamiltonian acts by means of the scalar ordinary Schr$\ddot{\rm o}$dinger operators:
\[
\hat{H}_0 = \frac{1}{2m}
\left[ -\hbar^2 \bigtriangleup +  \frac{\partial W}{\partial q^\mu} \, g^{\mu\mu} \frac{\partial W}{\partial q^\mu}  + \hbar
\bigtriangleup W \right] \  ,\quad \hat{H}_2 = \frac{1}{2m} \left[ -\hbar^2 \bigtriangleup +
 \frac{\partial W}{\partial q^\mu} \, g^{\mu\mu} \frac{\partial W}{\partial q^\mu}   -
\hbar \bigtriangleup W\right]
\]
In the Fermionic sector, however, the super-Hamiltonian reduces to a $2\times 2$-matrix Schr$\ddot{\rm o}$dinger operator with diagonals components:
\begin{eqnarray*} \hat{H}_1^{(11)} &=&
\frac{1}{2m} \left[ -\hbar^2 \bigtriangleup +  \frac{\partial W}{\partial q^\mu} \, g^{\mu\mu} \frac{\partial W}{\partial q^\mu}  + \hbar^2 \left( e_1^1 \frac{\partial \omega_1}{\partial
q^1}-e_2^2 \frac{\partial \omega_2}{\partial q^2}\right) + \hbar
\Box  W \right]
\\ && \\
 \hat{H}_1^{(22)} &=& \frac{1}{2m} \left[ -\hbar^2 \bigtriangleup
+  \frac{\partial W}{\partial q^\mu} \, g^{\mu\mu} \frac{\partial W}{\partial q^\mu}  +\hbar^2 \left( e_1^1 \frac{\partial \omega_1}{\partial
q^1}-e_2^2 \frac{\partial \omega_2}{\partial q^2}\right) - \hbar
\Box W \right]
\end{eqnarray*}
and odd diagonal components:
\begin{eqnarray*} \hat{H}_1^{(12)} &=&
\frac{1}{2m} \left[ -\hbar^2  \left( e_1^1 \omega_2 - e_2^2
\frac{\partial e_1^1}{\partial q^2} \right) \frac{\partial}{\partial
q^1} -\hbar^2   \left( e_2^2 \omega_1 + e_1^1 \frac{\partial e_2^2}{\partial
q^1} \right) \frac{\partial}{\partial q^2}  - \hbar^2  \left( e_2^2 \frac{\partial \omega_1}{\partial
q^2} + e_1^1 \frac{\partial \omega_2}{\partial q^1} \right) \right.
\\ && \left. +\hbar \left( \left( e_1^1 \omega_2 - e_2^2
\frac{\partial e_1^1}{\partial q^2} \right) \frac{\partial
W}{\partial q^1} - \left( e_2^2 \omega_1 + e_1^1 \frac{\partial
e_2^2}{\partial q^1} \right) \frac{\partial W}{\partial q^2}- e_1^1
e_2^2 \frac{\partial^2 W}{\partial q^1 \partial q^2} \right) \right]
\\
 \hat{H}_1^{(21)} &=&
\frac{1}{2m} \left[ \hbar^2  \left( e_1^1 \omega_2 - e_2^2
\frac{\partial e_1^1}{\partial q^2} \right) \frac{\partial}{\partial
q^1} +  \hbar^2  \left( e_2^2 \omega_1 + e_1^1 \frac{\partial e_2^2}{\partial
q^1} \right) \frac{\partial}{\partial q^2} + \hbar^2  \left( e_2^2 \frac{\partial \omega_1}{\partial
q^2} + e_1^1 \frac{\partial \omega_2}{\partial q^1} \right) \right.
\\ && \left. +\hbar \left( \left( e_1^1 \omega_2 - e_2^2
\frac{\partial e_1^1}{\partial q^2} \right) \frac{\partial
W}{\partial q^1} - \left( e_2^2 \omega_1 + e_1^1 \frac{\partial
e_2^2}{\partial q^1} \right) \frac{\partial W}{\partial q^2}- e_1^1
e_2^2 \frac{\partial^2 W}{\partial q^1 \partial q^2} \right) \right]
\end{eqnarray*}

\subsection{Systems separable in elliptic coordinates: SUSY QM}

We first analyze the case of separability in elliptic coordinates. The Euler version of these coordinates is defined as:
\begin{equation}
x^1\, =\, \frac{1}{d}\, u\, v\quad,\qquad x^2\, =\,\pm  \frac{1}{d} \, \sqrt{u^2-d^2}\, \sqrt{d^2-v^2}\label{ellipticeuler}
\end{equation}
and the inverse change is determined by: $u\, =\, \frac{r_1+r_2}{2}$, $v\, =\, \frac{r_2-r_1}{2}$. $r_1$ and $r_2$ denote the distances: $r_1\, =\, \sqrt{(x_1-d)^2+x_2^2}$\, , $r_2\, =\, \sqrt{(x_1+d)^2+x_2^2}$, from a generic point $(x^1,x^2)$ to two fixed points: $(d,0)$ and $(-d,0)$, the foci of the change of coordinates. Elliptic coordinates are restricted to the ranges: $d\,<\, u\, <\, \infty $, $-d\, <\, v\, <\, d$. Thus this change applies the cartesian plane ${\mathbb R}^2$, with coordinates $(x^1,x^2)$, into the elliptic strip ${\mathbb M}^2$ with coordinates $q^1\equiv u$, $q^2\equiv v$. The metric tensor is:
\[
g_{uu}\, =\, \frac{u^2-v^2}{u^2-d^2} \  ,\quad g_{vv}\, =\, \frac{u^2-v^2}{d^2-v^2}\, ,\qquad g_{uv}=g_{vu}=0
\]
The bosonic operators in elliptic coordinates are written as:
\[ \hat{p}_u =-i\hbar\frac{\partial}{\partial u}\,  ,
\, \hat{u}=u \, ,\quad \hat{p}_{v} =-i\hbar\frac{\partial}{\partial v}\,  , \,
\,  \hat{v}=v \quad  , \quad
[\hat{u},\hat{p}_u]= [\hat{v},\hat{p}_{v}]=i\hbar
\]
and the natural zweibein:
\[ e_1^{u} = \sqrt{\frac{u^2-d^2}{u^2-v^2}} \ , \
e_2^u = 0 \ ,\ e_1^v = 0 \ , \  e_2^{v}=
\sqrt{\frac{d^2-v^2}{u^2-v^2}}
\]
determines the Fermi operators: $\hat{\psi}^{u\dag} = e_1^u
\hat{\psi}^{1\dag}$, $\hat{\psi}^{u} = e_1^u \hat{\psi}^{1}$, $\hat{\psi}^{v\dag} = e_2^{v} \hat{\psi}^{2\dag} $, $\hat{\psi}^{v} = e_2^{v} \hat{\psi}^{2}$. Finally the supercharges are expressed as
\begin{equation}
\hat{Q}_E=\frac{i}{\sqrt{m}}\left(
\begin{array}{cccc}
 0 & D_u + \frac{\hbar u }{u^2-v^2} e_1^u & D_{v}-\frac{\hbar v}{u^2-v^2} e_2^v & 0 \\
 0 & 0 & 0 & \hspace{-0.2cm} -D_{v}  \\
 0 & 0 & 0 & D_u  \\
 0 & 0 & 0 & 0 \\
 \end{array}
\right) \label{qpe}
 \end{equation}
 \begin{equation}
 \hat{Q}_E^\dagger=\frac{i}{\sqrt{m}}\left(
\begin{array}{cccc}
 0 & 0 & 0 & 0  \\
\bar{D}_u & 0 & 0 & 0 \\
 \bar{D}_{v} & 0 & 0 & 0  \\
 0 & \hspace{-0.2cm} -\bar{D}_{v} + \frac{\hbar v}{u^2-v^2} e_2^v & \bar{D}_u + \frac{\hbar u}{u^2-v^2} e_1^u & 0  \\
 \end{array}
\right) \label{qmpe}
\end{equation}
and allow to construct the super-Hamiltonian: $\hat{H}_E=\frac{1}{2}\{\hat{Q}_E,\hat{Q}_E^\dagger\}$:
\begin{eqnarray*} \hat{H}_{E0} &=&
\frac{1}{2m} \left[-\hbar^2 \bigtriangleup + \left( \frac{u^2-d^2}{u^2-v^2}
\left(\frac{\partial W}{\partial u}\right)^2 + \frac{d^2-v^2}{u^2-v^2}
 \left(\frac{\partial W}{\partial v}\right)^2 \right) + \hbar \bigtriangleup W \right] \\ && \\
\hat{H}_{E2}&=& \frac{1}{2m}\left[-\hbar^2 \bigtriangleup +  \left( \frac{u^2-d^2}{u^2-v^2}
\left(\frac{\partial W}{\partial u}\right)^2 + \frac{d^2-v^2}{u^2-v^2}
 \left(\frac{\partial W}{\partial v}\right)^2 \right) - \hbar
\bigtriangleup W \right]
\\ &&\\ \hat{H}_{E1}^{(11)} &=&
\frac{1}{2m} \left[-\hbar^2 \bigtriangleup + \left( \frac{u^2-d^2}{u^2-v^2}
\left(\frac{\partial W}{\partial u}\right)^2 + \frac{d^2-v^2}{u^2-v^2}
 \left(\frac{\partial W}{\partial v}\right)^2 \right)+ \hbar^2
\frac{u^2+v^2-d^2}{(u^2-v^2)^2} - \hbar \Box W \right] \\ & & \\
\hat{H}_{E1}^{(22)} &=& \frac{1}{2m} \left[-\hbar^2 \bigtriangleup
+ \left( \frac{u^2-d^2}{u^2-v^2}
\left(\frac{\partial W}{\partial u}\right)^2 + \frac{d^2-v^2}{u^2-v^2}
 \left(\frac{\partial W}{\partial v}\right)^2 \right)+
 \hbar^2 \frac{u^2+v^2-d^2}{(u^2-v^2)^2} + \hbar \Box W \right]
\\ &&\\ \hat{H}_{E1}^{(12)} &=&
\frac{\hbar \sqrt{(u^2-d^2)(d^2-v^2)} }{m (u^2-v^2)^2} \left( \hbar u \frac{\partial}{\partial v}
+ \hbar   v \frac{\partial}{\partial u}  +  u
\frac{\partial W}{\partial v} - v \frac{\partial W}{\partial u} -
(u^2-v^2) \frac{\partial^2 W}{\partial u \partial
v} \right) \\ && \\
\hat{H}_{E1}^{(21)} &=& \frac{\hbar \sqrt{(u^2-d^2)(d^2-v^2)} }{m (u^2-v^2)^2} \left( -\hbar u \frac{\partial}{\partial v}
- \hbar   v \frac{\partial}{\partial u}  +  u
\frac{\partial W}{\partial v} - v \frac{\partial W}{\partial u} -
(u^2-v^2) \frac{\partial^2 W}{\partial u \partial
v} \right)
\end{eqnarray*}

\subsection{Systems separable in polar coordinates: SUSY QM}

Reproducing the results for the case of standard polar coordinates: $x^1 = r\, \cos \varphi$, $x^2 = r \, \sin \varphi$,
and identifying: $q^1\equiv r$, $q^2\equiv \varphi$, the Bosonic operators read:
\[
\hat{p}_r =-i\hbar\frac{\partial}{\partial r}\,  ,
\quad  \hat{r}=r \,  , \quad \hat{p}_{\varphi} =-i\hbar\frac{\partial}{\partial \varphi}\,  , \,
\,  \hat{\varphi}=\varphi\, , \quad [\hat{r},\hat{p}_r]=[\hat{\varphi},\hat{p}_{\varphi}]=i\hbar
\]
The natural zweibein:  $ e_1^{r} = 1$, $e_2^r = 0$, $e_1^{\varphi} = 0$, $e_2^{\varphi}= \frac{1}{r}$, determine the Fermi operators: $\hat{\psi}^{r\dag} = e_1^r
\hat{\psi}^{1\dag}$, $ \hat{\psi}^r = e_1^r
\hat{\psi}^{1}$, $\hat{\psi}^{\varphi\dag} = e_2^{\varphi} \hat{\psi}^{2\dag}$ and $ \hat{\psi}^{\varphi} = e_2^{\varphi} \hat{\psi}^{2}$, and the supercharges are written in polar coordinates as follows:
\begin{equation}
\hat{Q}_P=\frac{i}{\sqrt{m}}\left(
\begin{array}{cccc}
 0 & D_r + \frac{\hbar}{r} e_1^r & D_{\varphi} & 0 \\
 0 & 0 & 0 & \hspace{-0.2cm} -D_{\varphi}  \\
 0 & 0 & 0 & D_r  \\
 0 & 0 & 0 & 0 \\
 \end{array}
\right) \,,\   \hat{Q}_P^\dagger=\frac{i}{\sqrt{m}}\left(
\begin{array}{cccc}
 0 & 0 & 0 & 0  \\
\bar{D}_r & 0 & 0 & 0 \\
 \bar{D}_{\varphi} & 0 & 0 & 0  \\
 0 & \hspace{-0.2cm} -\bar{D}_{\varphi} & \bar{D}_r + \frac{\hbar}{r} e_1^r & 0  \\
 \end{array}
\right) \label{qmp}
\end{equation}
The general super-Hamiltonian is:
\begin{eqnarray*} \hat{H}_{P0} &=&
\frac{1}{2m} \left[-\hbar^2 \bigtriangleup + \left(
\left(\frac{\partial W}{\partial r}\right)^2 +
\frac{1}{r^2} \left(\frac{\partial W}{\partial \varphi}\right)^2 \right) +\hbar \bigtriangleup W\right] \\
\hat{H}_{P2}&=& \frac{1}{2m}\left[-\hbar^2 \bigtriangleup + \
\left( \left(\frac{\partial W}{\partial r}\right)^2 + \frac{1}{r^2}
\left(\frac{\partial W}{\partial \varphi}\right)^2 \right) - \hbar
\bigtriangleup W \right] \\ \hat{H}_{P1}^{(11)} &=&
\frac{1}{2m} \left[-\hbar^2 \bigtriangleup + \left(
\left(\frac{\partial W}{\partial r}\right)^2 + \frac{1}{r^2}
\left(\frac{\partial W}{\partial \varphi}\right)^2 \right) + \hbar^2
\frac{1}{r^2} - \hbar \Box W \right] \\
\hat{H}_{P1}^{(22)} &=& \frac{1}{2m} \left[-\hbar^2 \bigtriangleup
+ \left( \left(\frac{\partial W}{\partial r}\right)^2 +
\frac{1}{r^2} \left(\frac{\partial W}{\partial \varphi}\right)^2
\right) + \hbar^2 \frac{1}{r^2} + \hbar \Box W \right]\\
\hat{H}_{P1}^{(12)} &=&
\frac{\hbar}{m r^2} \left( \hbar \frac{\partial}{\partial
\varphi}  + \frac{\partial W}{\partial
\varphi} - \frac{r^2}{2} \frac{\partial^2 W}{\partial r \partial
\varphi} \right)\  , \
\hat{H}_{P1}^{(21)} =\frac{\hbar}{m r^2} \left( -\hbar \frac{\partial}{\partial
\varphi}  + \frac{\partial W}{\partial
\varphi} - \frac{r^2}{2} \frac{\partial^2 W}{\partial r \partial
\varphi} \right)
\end{eqnarray*}

\subsection{Systems separable in parabolic coordinates: SUSY QM}

Finally, we present the results corresponding to parabolic coordinates: $q^1\equiv \xi_1$, $q^2 \equiv \xi_2$. The change of coordinates is given by:  $x^1\, =\, \frac{1}{2} \, \left( \xi_1^2- \xi_2^2 \right)$, $x^2\, =\, \xi_1 \, \xi_2$, with $-\infty\, <\, \xi_1\, <\, \infty$, and $0\, \leq \, \xi_2\,
<\, \infty$. The metric tensor is: $g_{11}=g_{22}=\xi_1^2+\xi_2^2$, $g_{12} =g_{21}= 0$, with a natural zweibein: $e_1^{\xi_1} = \frac{1}{\sqrt{\xi_1^2 + \xi_2^2}}$, $e_2^{\xi_1}=e_1^{\xi_2} = 0$, $e_2^{\xi_2} = \frac{1}{\sqrt{\xi_1^2 + \xi_2^2}}$. Thus the supercharges in parabolic coordinates are:
\begin{equation}
\hat{Q}_{{\cal P}}=\frac{i}{\sqrt{m}}\left(
\begin{array}{cccc}
 0 & D_{\xi_1} + \frac{\hbar \xi_1}{\xi_1^2+\xi_2^2} e_1^{\xi_1} &
 D_{\xi_2} + \frac{\hbar \xi_2}{\xi_1^2+\xi_2^2} e_2^{\xi_2} & 0 \\
 0 & 0 & 0 & \hspace{-0.2cm} -D_{\xi_2}  \\
 0 & 0 & 0 & D_{\xi_1}  \\
 0 & 0 & 0 & 0 \\
 \end{array}
\right) \label{qpb}
 \end{equation}
\begin{equation}
 \hat{Q}_{{\cal P}}^\dagger=\frac{i}{\sqrt{m}}\left(
\begin{array}{cccc}
 0 & 0 & 0 & 0  \\
\bar{D}_{\xi_1} & 0 & 0 & 0 \\
 \bar{D}_{\xi_2} & 0 & 0 & 0  \\
 0 & \hspace{-0.2cm} -\bar{D}_{\xi_2} -\frac{\hbar \xi_2}{\xi_1^2+\xi_2^2} e_2^{\xi_2} &
 \bar{D}_{\xi_1} + \frac{\hbar \xi_1}{\xi_1^2+\xi_2^2} e_1^{\xi_1}  & 0  \\
 \end{array}
\right) \label{qmpb}
\end{equation}
And the super-Hamiltonian is:
\begin{eqnarray*} \hat{H}_{{\cal P}0} &=&
\frac{1}{2m} \left[-\hbar^2 \bigtriangleup +
\frac{1}{\xi_1^2+\xi_2^2} \left( \left(\frac{\partial W}{\partial
\xi_1}\right)^2 +
 \left(\frac{\partial W}{\partial \xi_2}\right)^2 \right) + \hbar \bigtriangleup W\right] \\
\hat{H}_{{\cal P}2} &=& \frac{1}{2m} \left[-\hbar^2 \bigtriangleup +
\frac{1}{\xi_1^2+\xi_2^2} \left( \left(\frac{\partial W}{\partial
\xi_1}\right)^2 + \left(\frac{\partial W}{\partial \xi_2}\right)^2
\right) - \hbar \bigtriangleup W\right]
\\  \hat{H}_{{\cal P}1}^{(11)} &=&
\frac{1}{2m} \left[-\hbar^2 \bigtriangleup +
\frac{1}{\xi_1^2+\xi_2^2} \left( \left(\frac{\partial W}{\partial
\xi_1}\right)^2 + \left(\frac{\partial W}{\partial \xi_2}\right)^2
\right)  + \frac{\hbar^2}{(\xi_1^2+\xi_2^2)^2} - \hbar \Box W \right] \\
 \hat{H}_{{\cal P}1}^{(22)} &=&
\frac{1}{2m} \left[-\hbar^2 \bigtriangleup +
\frac{1}{\xi_1^2+\xi_2^2} \left( \left(\frac{\partial W}{\partial
\xi_1}\right)^2 + \left(\frac{\partial W}{\partial \xi_2}\right)^2
\right) + \frac{\hbar^2}{(\xi_1^2+\xi_2^2)^2} + \hbar \Box W \right]
\\ \hat{H}_{{\cal P}1}^{(12)} &=&
\frac{\hbar}{m (\xi_1^2+\xi_2^2)^2} \left( \hbar  \xi_1
\frac{\partial}{\partial \xi_2} -  \hbar \xi_2
\frac{\partial}{\partial \xi_1}  +
 \xi_2
\frac{\partial W}{\partial \xi_1}+ \xi_1
\frac{\partial W}{\partial \xi_2} - (\xi_1^2+\xi_2^2) \frac{\partial^2 W}{\partial
\xi_1 \partial \xi_2} \right)
\\ \hat{H}_{{\cal P}1}^{(21)} &=& \frac{\hbar}{m (\xi_1^2+\xi_2^2)^2} \left( -\hbar  \xi_1
\frac{\partial}{\partial \xi_2} +  \hbar \xi_2
\frac{\partial}{\partial \xi_1}  +
 \xi_2
\frac{\partial W}{\partial \xi_1}+ \xi_1
\frac{\partial W}{\partial \xi_2} - (\xi_1^2+\xi_2^2) \frac{\partial^2 W}{\partial
\xi_1 \partial \xi_2} \right)
\end{eqnarray*}

\section{The SUSY two-fixed Coulomb centers problem}

We present as an example the supersymmetric problem of two Coulombian fixed centers, determined by the quantum Hamiltonian:
\[
\hat{H}=\, -\frac{\hbar^2 }{2 m} \bigtriangleup - \frac{\alpha_1}{r_1}
- \frac{\alpha_2}{r_2}
\]
where the two centers are located at the points $(-d,0)$ and $(d,0)$. $\alpha_1$ and $\alpha_2$ denote the center strengths, and we assume, without loss of generality: $\alpha_1=\alpha \geq \alpha_2=\delta\alpha$, with $\delta \in (0,1]$. Quantization of the charge implies that: $\alpha=e^2 Z_1$, $e$ being the charge of the electron, and $\delta=\frac{Z_2}{Z_1} \leq 1$, with $Z_1,Z_2\in {\mathbb N}^*$, $Z_1\geq Z_2$.

In order to construct a supersymmetric version of this problem, we choose the superpotential of the system among the solutions of the Poisson equation:
\begin{equation}
 {\hbar \over 2 m } \nabla^2 W = V\label{eq:poiss1}
\end{equation}
in a similar way to what has been done in references \cite{Ioffe1} and \cite{Wipf, Wipf2} for the Kepler/Coulomb problem. We will follow the notation introduced in \cite{Sigma}. Equation (\ref{eq:poiss1}) is separable in elliptic coordinates. Considering the ansatz: $W(u,v)=F(u)+G(v)$, the equation is written as:
\[
{{\hbar} \over 2 m} \left[ {u^2 - d^2 \over u^2-v^2} \left( {d^2
F \over d u^2} + {u\over u^2-d^2 }{d F \over d u} \right)
+ {d^2 -v^2 \over u^2-v^2} \left( {d^2 G \over d v^2}-{v\over
d^2-v^2} {d G \over d v} \right) \right] = -{ \alpha (1+\delta) u \over
u^2-v^2} - { \alpha (1- \delta)  v\over u^2-v^2}
\]
that reduces to a ODE system solved by the general expressions:
\begin{eqnarray*}
F(u)&=& -  \frac{2 m \alpha (1+\delta)}{{\hbar}} u + {\kappa \over
2} \left( {\rm ln} \left(\frac{u}{d}+ \frac{1}{d} \sqrt{u^2-d^2}\right) \right)^2 + C_1 {\rm ln} \left(\frac{u}{d}+\frac{1}{d} \sqrt{u^2-d^2}\right) + C_3  \\  G(v)&=&  \frac{ 2 m \alpha (1-\delta)}{{\hbar}} v - {\kappa
\over 2} \left( {\rm arcsin}\frac{v}{d} \right)^2  + C_2 {\rm arcsin}\ \frac{v}{d}  +
C_4
\end{eqnarray*}
We shall restrict ourselves to the the simplest case where the separation constant $\kappa$ and the integration constants are zero: $\kappa=C_1=C_2=C_3=C_4=0$,
\begin{equation}
W(u,v)\, =\,   - \frac{2 m \alpha}{\hbar} \left(  (1+\delta) u-
(1-\delta) v\right)  \label{eq:sups}
\end{equation}
and thus expressions (\ref{qpe}) and (\ref{qmpe}) for supercharges in elliptic coordinates become:
\[ D_u=e_1^u \left( \hbar \frac{\partial}{\partial u}
- \frac{2 m \alpha (1+\delta)}{\hbar}\right) \,  , \,
D_{v}=e_2^{v} \left( \hbar
\frac{\partial}{\partial v} + \frac{2 m \alpha (1-\delta)}{\hbar}\right)
\]
\[
\bar{D}_u= e_1^u \left( \hbar
\frac{\partial}{\partial u}+\frac{2 m \alpha (1+\delta)}{\hbar}\right)\quad \, , \,  \bar{D}_{v}= e_2^{\varphi} \left(
\hbar \frac{\partial}{\partial v}-\frac{2 m \alpha (1-\delta)}{\hbar}\right)
\]
that, in turn, lead to the super-Hamiltonian of the system:
\begin{eqnarray*}
\hat{H}_{E0} &=&
\frac{1}{2m} \left[-\hbar^2 \bigtriangleup + \frac{ 4 m \alpha^2}{\hbar^2} \, \frac{ (1+\delta)^2 u^2 -
(1-\delta)^2 v^2 - 4 d^2 \delta}{u^2-v^2}-\frac{2 \alpha (1+\delta) u + 2 \alpha (1-\delta) v}{u^2-v^2} \right]\\ \hat{H}_{E2} &=&
\frac{1}{2m} \left[-\hbar^2 \bigtriangleup + \frac{ 4 m \alpha^2}{\hbar^2} \, \frac{ (1+\delta)^2 u^2 -
(1-\delta)^2 v^2 - 4 d^2 \delta}{u^2-v^2}+\frac{2 \alpha (1+\delta) u + 2 \alpha (1-\delta) v}{u^2-v^2} \right]
\end{eqnarray*}
\begin{eqnarray*} \hat{H}_{E1}^{(11)} &=&
\frac{1}{2m} \left[-\hbar^2 \bigtriangleup + \frac{4 m \alpha^2}{\hbar^2} \frac{ ((1+\delta)^2 u^2 - (1-\delta)^2 v^2 - 4 d^2 \delta)}{u^2-v^2} +\hbar^2
\frac{u^2+v^2-d^2}{(u^2-v^2)^2} \right. \\ &-& \left.   \frac{ (\alpha (1+ \delta) u + \alpha (1-\delta) v) (u^2+v^2 - 2 d^2)}{(u^2-v^2)^2} \right]\\
\hat{H}_{E1}^{(22)} &=& \frac{1}{2m} \left[-\hbar^2 \bigtriangleup
+ \frac{4 m \alpha^2}{\hbar^2} \frac{ ((1+\delta)^2 u^2 - (1-\delta)^2 v^2 - 4 d^2 \delta)}{u^2-v^2}  + \hbar^2 \frac{u^2+v^2-d^2}{(u^2-v^2)^2} \right. \\ &+& \left.
 \frac{ (\alpha (1+ \delta) u + \alpha (1-\delta) v) (u^2+v^2 - 2 d^2)}{(u^2-v^2)^2}\right]\\
 \hat{H}_{E1}^{(12)} &=& \frac{\sqrt{(u^2-d^2)(d^2-v^2)}}{(u^2-v^2)^2} \left(
\frac{\hbar^2}{m} \left( u \frac{\partial}{\partial
v}  +   v\frac{\partial}{\partial
u} \right) +  2 \alpha (1-\delta) u + 2 \alpha (1+\delta) v \right)\\
\hat{H}_{E1}^{(21)} &=&
\frac{\sqrt{(u^2-d^2)(d^2-v^2)}}{(u^2-v^2)^2} \left(
\frac{-\hbar^2}{m} \left( u \frac{\partial}{\partial
v}  +   v\frac{\partial}{\partial
u} \right) +  2 \alpha (1-\delta) u + 2 \alpha (1+\delta) v \right)
\end{eqnarray*}
In \cite{CM} we have found many bound state eigenfunctions of $\hat{H}_{E0}$ profiting from the separability of the corresponding Schr\"odinger equation in elliptic coordinates. Also, the partner eigenfunctions in the discrete spectrum of $\hat{H}_{E1}$ have been constructed through the action of $\hat{Q}^{\dag}$. We will present here only the specific results for the zero modes of the system. The zero energy wave functions for the scalar Hamiltonians can easily be found as those annihilated by the two supercharges: $\hat{Q}_E \Psi_0^{(0)}(u, v)=0$, $\hat{Q}_E^{\dag} \Psi_0^{(0)}(u ,v)=0$; $\hat{Q}_E \Psi_0^{(2)}(u, v)=0$, $\hat{Q}_E^{\dag} \Psi_0^{(2)}(u ,v)=0$. Solving the corresponding differential equations, the solutions are:
\begin{equation}
\Psi_0^{(0)}(u,v)= A_1 \left(
\begin{array}{c} e^{- \frac{2 m \alpha (1+\delta) u}{\hbar^2} + \frac{ 2 m \alpha
(1-\delta) v}{\hbar^2}}
\\ 0 \\ 0 \\ 0
\end{array} \right)\  ,\quad  \Psi_0^{(2)}(u,v)=   A_2 \left(
\begin{array}{c} 0
\\ 0 \\ 0 \\ e^{\frac{2 m \alpha (1+\delta) u}{\hbar^2} - \frac{ 2 m \alpha
(1-\delta) v}{\hbar^2}}
\end{array} \right)
\end{equation}
Only $\Psi_0^{(0)}(u,v)$ is of finite norm:
\begin{eqnarray*}
N^2&=& 2\pi d \left[
\frac{\hbar^2}{4 m \alpha (1+\delta)} I_0\left(\frac{4 m \alpha\, d
(1-\delta) }{\hbar^2}\right)\, K_1\left(\frac{4 m \alpha\, d
(1+\delta) }{\hbar^2}\right) \right. \\ && \left. +\frac{\hbar^2}{4 m \alpha (1-\delta)}
I_1\left( \frac{4 m \alpha\ d (1-\delta) }{\hbar^2}\right)\,
K_0\left(\frac{4 m \alpha\ d (1+\delta) }{\hbar^2}\right) \right]
\end{eqnarray*}
where $I_n$ and $K_n$ represent the modified Bessel functions. Henceforth $\Psi_0^{(0)}(u,v)$ is a ground state. We can write the zero modes in Cartesian coordinates: $\Psi_0^{(0)}(x^1,x^2)= S  \Psi_0^{(0)}(u,v)$, $\Psi_0^{(2)}(x^1,x^2)=S  \Psi_0^{(2)}(u,v)$,  where $S$ denotes the matrix of the change of coordinates in the spinor space:
\[ S = \left( \begin{array}{cccc} 1 & 0 & 0 & 0 \\
0 &{ \frac{v}{d}} e_1^u & {\frac{u}{d}} e_2^v & 0 \\ 0 & {
\frac{u}{d}} e_2^v & - {  \frac{v}{d}} e_1^u & 0
\\ 0 & 0 & 0 & -1 \end{array} \right)
\]
Thus:
\[
\Psi_0^{(0)}(x^1,x^2)= A_1 \left(
\begin{array}{c} e^{{-2 m \alpha r_1-2 m \alpha \delta r_2 \over \hbar^2}}
\\ 0 \\ 0 \\ 0
\end{array} \right)\, ,\quad  \Psi_0^{(2)}(x^1,x^2) =  A_2 \left(
\begin{array}{c} 0
\\ 0 \\ 0 \\ - e^{{2 m \alpha r_1+2 m \alpha \delta r_2 \over \hbar^2}}
\end{array} \right)
\]

The Fermionic zero modes are annihilated by the two supercharges:
\[
\hat{Q}_E \Psi_0^{(1)}(u,v)=0 \ , \
\hat{Q}_E^{\dag} \Psi_0^{(1)}(u,v)=0 \ , \
\Psi_0^{(1)}(u,v)= \left(\begin{array}{c} 0 \\
\psi^{(1)1}_0(u,v) \\ \psi^{(1)2}_0(u,v) \\ 0
\end{array}\right)
\]
and the resolution of this system of differential equations gives us two independent solutions:
\[
\Psi_0^{(1)}(u,v)=A_1 \frac{1}{\sqrt{u^2-v^2}}\left(
\begin{array}{c} 0\\
e^{ \frac{ 2 m \alpha (1+\delta) u + 2 m \alpha (1-\delta) v}{\hbar^2}} \\ 0 \\ 0
\end{array} \right) + A_2 \frac{1}{\sqrt{u^2-v^2}}\left(
\begin{array}{c} 0\\0\\
e^{- \frac{ 2 m \alpha (1+\delta) u + 2 m \alpha (1-\delta) v}{\bar{\hbar}^2}} \\ 0
\end{array} \right)
\]
Only the second one is normalizable, thus taking $A_1=0$ we find the Fermionic ground state:
\[
\Psi_0^{(1)}(u,v)=  A_2 \frac{1}{\sqrt{u^2-v^2}}\left(
\begin{array}{c} 0\\ 0
\\e^{- \frac{ 2 m \alpha (1+\delta) u + 2 m \alpha (1-\delta) v}{\bar{\hbar}^2}}   \\ 0
\end{array} \right)
\]
with norm:
\[
N^2 =\, 2 \pi K_0 \left( {4 m \alpha\, d   \over
\hbar^2} (1+\delta) \right) I_0 \left( {4 m \alpha\, d \over
\hbar^2} (1-\delta) \right)
\]
In Cartesian coordinates, $\Psi_0^{(1)}(x^1,x^2)= S
\Psi^{(1)}_0(u,v)$, we find:
\begin{equation}
\Psi_0^{(1)}(x^1,x^2) = \frac{1}{N} \left(
\begin{array}{c} 0\\
- \frac{(r_1+r_2)}{4 d \sqrt{r_1r_2}} \sqrt{\frac{4 d^2}{r_1 r_2}
-\frac{r_1}{r_2}-\frac{r_2}{r_1}+2}
\\  \frac{(r_2-r_1)}{4 d \sqrt{r_1r_2}} \sqrt{-\frac{4  d^2}{r_1 r_2}
+\frac{r_1}{r_2}+\frac{r_2}{r_1}+ 2} \\ 0
\end{array} \right) e^{-\frac{2 m \alpha ( \delta r_1 +r_2)}{\hbar^2}}
\end{equation}

\subsection{The SUSY Kepler/Coulomb problems Limit}

There exists two natural ways to obtain the Kepler/Coulomb problem from the problem of two centers: 1. The two centers collapse in one, and 2. One center flies to infinity. Taking into account that polar and parabolical coordinates can be regarded as limiting cases of elliptic coordinates, we expect that the SUSY Kepler/Coulomb problem either in the polar separable version (Limit 1) or, in the parabolic separable version (Limit 2) will arise in these limits of the elliptic system of coordinates. If this is the case, a path to the explanation of the superintegrability of the Kepler/Coulomb problem will be open.

Following \cite{Ioffe1} and \cite{Wipf} we consider as superpotential and super-Hamiltonian for the Kepler/Coulomb problem the following expressions:
\[
W(r,\varphi) =- \frac{2}{\hbar} r\, ,\quad \hat{H}_{P0}=
-\frac{\hbar^2}{2}\bigtriangleup + \frac{2}{\hbar^2} -\frac{1}{r}\,
,\quad \hat{H}_{P2}= -\frac{\hbar^2}{2}\bigtriangleup +
\frac{2}{\hbar^2} +\frac{1}{r}
\]
for the bosonic sectors (scaled variables are used), and
\[
\hat{H}_{P1}=\left(\begin{array}{ccc}
-\frac{\hbar^2}{2}\bigtriangleup + \frac{2}{\hbar^2} +
\frac{\hbar^2}{2 r^2} - \frac{1}{r} & \frac{\hbar^2}{r^2}
\frac{\partial}{\partial \varphi}
\\ & \\ -\frac{\hbar^2}{r^2} \frac{\partial}{\partial \varphi}&
-\frac{\hbar^2}{2}\bigtriangleup +
\frac{2}{\hbar^2}+\frac{\hbar^2}{2 r^2}+ \frac{1}{r}
\end{array}\right) \quad .
\]
for the $2\times 2$ Fermionic block.

\subsubsection{The two centers collapse into one center.}

We will use the trigonometric version of the elliptic coordinates in order to guarantee the preservation of the orientation:
$u\, =\, \cosh (\xi -\beta)$, $v=\cos \eta$.  $\eta \in [ 0 , 2 \pi]$, and $\beta\leq \xi <\infty$ is an arbitrary constant.
\[
\left\{ \begin{array}{l} x^1 = d \cosh (\xi-\beta) \cos\eta \quad \qquad , \quad \qquad
 x^2 = d \sinh (\xi-\beta) \sin \eta \end{array} \right. \quad .
\]
The limit from elliptic to polar coordinates is obtained by taking $d = 2 e^{\beta}$ and the arbitrary constant $\beta \rightarrow -\infty$. This limit produces $d\to 0$ and thus the two centers will collapse into only one.
\[ x^1 = \lim_{\beta
\rightarrow -\infty} d \cosh (\xi-\beta)
\cos\eta = r \cos\varphi \, ,\quad
x^2 = \lim_{\beta \rightarrow -\infty} d \sinh (\xi-\beta)
\sin\eta = r \sin\varphi
\]
where $e^{\xi}\equiv r$. Applying this limits to the supercharges and the supersymmetric
Hamiltonian of the two-center Euler/Pauli problem we obtain in a direct way the corresponding expressions  for the Kepler/Coulomb problem:
\[ \hat{Q}_P = \lim_{d \rightarrow 0}
\hat{Q}_E \ , \ \hat{Q}_P^{\dag} = \lim_{d \rightarrow 0}
\hat{Q}_E^{\dag} \ , \ \hat{H}_P =  \lim_{d \rightarrow 0} \hat{H}_E\, ,\quad
W(r, \varphi)\, =\,  \lim_{d\rightarrow 0} W(\xi,\eta)\, =\,  - \frac{2
m \tilde{\alpha}  }{{\hbar}} r
\]
with $\tilde{\alpha} = \alpha (1+\delta)$, i.e., the electric charge is the sum of the corresponding charges of the two centers.

We have shown in \cite{CM} that eigenfunctions in the discrete spectrum of the bosonic super-Hamiltonian of the two center problem go smoothly to all the normalizable eigenfunctions in the Kepler/Coulomb system when the two centers collapse. The zero modes, however, behave in a different way such that an interesting situation arises. Taking the limit polar limit in the bosonic zero modes and its norm we find:
\[ \Psi_0^{(0)}(r ,\varphi) =
\lim_{d\rightarrow 0} \Psi_0^{(0)}(\xi ,\eta)= \frac{2 m
\tilde{\alpha}}{\hbar^2} \sqrt{\frac{2}{\pi}} \left(
\begin{array}{c} e^{- \frac{2 m \tilde{\alpha}}{\hbar^2} r}
\\ 0 \\ 0 \\ 0
\end{array} \right) \qquad , \qquad
\lim_{d\rightarrow 0}  N^2  = \frac{\pi \hbar^4}{8 m^2
\tilde{\alpha}^2 } \quad .
\]
Thus, we check that the well-known bosonic zero mode of the SUSY Kepler/Coulomb problem expressed in polar coordinates arises in this limit.

Simili modo, the Fermionic zero energy ground state of the SUSY Euler/Pauli problem becomes in this limit $d\rightarrow 0$:
\[
\Psi_0^{(1)}(r,\varphi) = \lim_{d\rightarrow
0} \Psi_0^{(1)}(u,v) =  A_2 \, \frac{1}{r}\left(
\begin{array}{c} 0\\0\\
e^{- \frac{ 2 m \tilde{\alpha}}{{\hbar}^2} r} \\
0 \end{array} \right)
\]
but the corresponding norm $N$ diverges:
\[ \lim_{d\rightarrow 0} N^2 =
\lim_{d\rightarrow 0} 2 \pi K_0 \left( {4 m \alpha\, d \over
\hbar^2} (1+\delta) \right) I_0 \left( {4 m \alpha\, d \over
\hbar^2} (1-\delta) \right) = \infty
\]
There is no  normalizable Fermionic zero modes in the SUSY Kepler/Coulomb system.

\subsubsection{One center \lq\lq escapes'' to infinity.}

Using re-scaled parabolic coordinates, the Ke\-pler/Cou\-lomb superpotential and super Hamiltonian for the Kepler/Coulomb system read:
\begin{eqnarray*}
W(\xi_1,\xi_2)& =&-\frac{1}{\hbar} (\xi_1^2+\xi_2^2)\\
\hat{H}_{{\cal P}0}&=&-\frac{\hbar^2}{2 (\xi_1^2+\xi_2^2)}\left(
\frac{\partial^2}{\partial \xi_1^2} + \frac{\partial^2}{\partial
\xi_2^2} \right) + \frac{2}{\hbar^2} -\frac{2}{\xi_1^2+\xi_2^2} \\
\hat{H}_{{\cal P}2}&=&-\frac{\hbar^2}{2 (\xi_1^2+\xi_2^2)}\left(
\frac{\partial^2}{\partial \xi_1^2} + \frac{\partial^2}{\partial
\xi_2^2} \right)
+ \frac{2}{\hbar^2} +\frac{2}{\xi_1^2+\xi_2^2} \\
\hat{H}_{{\cal P}1}^{(11)}&=& -\frac{\hbar^2}{2
(\xi_1^2+\xi_2^2)}\left( \frac{\partial^2}{\partial \xi_1^2} +
\frac{\partial^2}{\partial \xi_2^2} \right)
+ \frac{2}{\hbar^2} + \frac{\hbar^2}{2 (\xi_1^2+\xi_2^2)^2} - \frac{2(\xi_1^2-\xi_2^2)}{(\xi_1^2+\xi_2^2)^2} \\ && \\
\hat{H}_{{\cal P}1}^{(22)}&=& -\frac{\hbar^2}{2
(\xi_1^2+\xi_2^2)}\left( \frac{\partial^2}{\partial \xi_1^2} +
\frac{\partial^2}{\partial \xi_2^2} \right)
+ \frac{2}{\hbar^2} + \frac{\hbar^2}{2 (\xi_1^2+\xi_2^2)^2} + \frac{2(\xi_1^2-\xi_2^2)}{(\xi_1^2+\xi_2^2)^2} \\ && \\
\hat{H}_{{\cal P}1}^{(12)}&=&  \frac{\hbar^2}{(\xi_1^2+\xi_2^2)^2}
\left(\xi_1 \frac{\partial}{\partial \xi_2}-\xi_2 \frac{\partial}{\partial \xi_1} \right) - \frac{4 \xi_1 \xi_2}{(\xi_1^2+\xi_2^2)^2} \\ && \\
\hat{H}_{{\cal P}1}^{(21)}&=&  - \frac{\hbar^2}{(\xi_1^2+\xi_2^2)^2}
\left(\xi_1 \frac{\partial}{\partial \xi_2}-\xi_2
\frac{\partial}{\partial \xi_1} \right) - \frac{4 \xi_1
\xi_2}{(\xi_1^2+\xi_2^2)^2}
\end{eqnarray*}
The limit from elliptic to parabolic coordinates not only requires to take $d\to \infty$, but also a translation of the origin: $
x^1=x-d$, $x^2=y$, and a re-definition of the elliptic coordinates $(u,v)$ as follows:
\[
\xi_1\, =\, \frac{ \pm \sqrt{u^2-d^2}}{\sqrt{d}}\, \quad,\qquad
\xi_2\, =\, \frac{\sqrt{d^2-v^2}}{\sqrt{d}}\quad .
\]
In the trigonometric version, these changes are written as: $
\xi_1 = \pm \sqrt{d} \sinh\xi \ , \ \xi_2 = \sqrt{d} |\sin\eta|$. Now, in the $d\to \infty$ limit two-center operators are converted into the Kepler/Coulomb ones in parabolic coordinates:
\[ \hat{Q}_{{\cal P}} = \lim_{d \rightarrow \infty}
\hat{Q}_E \ , \ \hat{Q}_{{\cal P}}^{\dag} = \lim_{d \rightarrow
\infty} \hat{Q}_E^{\dag} \ , \ \hat{H}_{{\cal P}} =  \lim_{d
\rightarrow \infty} \hat{H}_E
\]
Moreover, the superpotential in this parabolic limit becomes the Euler/Coulomb superpotential in parabolic
coordinates:
\[
W(\xi_1, \xi_2) = \lim_{d\rightarrow \infty} W(\xi,\eta)\, =\,  - \frac{
m \alpha  }{{\hbar}} (\xi_1^2+\xi_2^2) \qquad .
\]
The Bosonic zero energy ground state of the SUSY
Euler/Pauli problem goes in this limit to:
\[ \Psi_0^{(0)}(\xi_1 ,\xi_2) =
\lim_{d\rightarrow \infty} \Psi_0^{(0)}(\xi ,\eta)= \frac{2 m
\alpha}{\hbar^2} \sqrt{\frac{2}{\pi}} \left(
\begin{array}{c} e^{- \frac{ m \alpha}{\hbar^2} (\xi_1^2+\xi_2^2)}
\\ 0 \\ 0 \\ 0
\end{array} \right) \, ,\quad
\lim_{d\rightarrow \infty}  N^2  = \frac{\pi \hbar^4}{8 m^2
\alpha^2 } \quad ,
\]
i.e., to the zero mode of the SUSY Kepler/Coulomb
problem in parabolic coordinates with electric charge $\alpha$.

The Fermionic zero energy ground state of
Euler/Coulomb problem goes to a zero energy eigenstate of the Kepler/Coulomb super Hamiltonian:
\[
\Psi_0^{(1)}(\xi_1,\xi_2) = \lim_{d\rightarrow
\infty} \Psi_0^{(1)}(\xi,\eta)
=  A_2 \, \frac{1}{\sqrt{\xi_1^2+\xi_2^2}}\left(
\begin{array}{c} 0\\0\\
e^{- \frac{  m \alpha}{{\hbar}^2} (\xi_1^2-\xi_2^2)} \\
0 \end{array} \right)
\]
but the norm of this state is again infinite. There are no normalizable Fermionic zero modes of the SUSY
Kepler problem both in the polar and parabolic regimes.


\section*{References}
\begin{thebibliography}{9}

\bibitem{Eisenhart} Eisenhart L P 1948 Enumeration of Potentials for Which One-Particle Schr\oe dinger Equations Are Separable {\it Phys. Rev.} {\bf 74} 87-89.


\bibitem{Ioffe} Andrianov A A, Ioffe M V and Tsu Z P 1988 The factorization method in curvilinear coordinates and level pairing for matrix potentials {\it Vestn. Leningrad Univ., Fiz. 4}  {\bf 25} 3-9 (in Russian), ({\it Preprint} arXiv:1101.0773 (in English)).

\bibitem{AoP} Alonso Izquierdo A, Gonzalez Leon M A, Mateos Guilarte J and de la Torre Mayado M 2003 Supersymmetry versus integrability in two-dimensional classical mechanics  {\it Ann. Phys.} {\bf 308} 664-691 ({\it Preprint} hep-th/0307123).

\bibitem{JPA} Alonso Izquierdo A, Gonzalez Leon M A, Mateos Guilarte J and de la Torre Mayado M 2004 On two-dimensional superpotentials: from classical Hamilton-Jacobi theory to 2D supersymmetric quantum mechanics {\it Journal of Physics A} {\bf 37} 10323-10338 ({\it Preprint} hep-th/0401054).

\bibitem{Sigma} Gonzalez Leon M A, Mateos Guilarte J and de la Torre M 2007  Two-dimensional Supersymmetric Quantum Mechanics: Two Fixed Centers of Force {\it SIGMA} {\bf 3} 124-148 ({\it Preprint}  arXiv:0712.3682).

\bibitem{CM} Gonzalez Leon M A, Mateos Guilarte J, de la Torre M and Senosiain M J 2011 On the Supersymmetric Spectra of two Planar Integrable Quantum Systems, To be published in {\it Cont. Math.} ({\it Preprint} arXiv:1107.4886).

\bibitem{Ioffe1} Andrianov A A, Borisov N and Ioffe M V 1984 Factorization method and the Darboux transformation for multidimensional hamiltonians {\it Theor. Math. Phys.} {\bf 61} 1078-1088.

\bibitem{Wipf} Kirchberg A, L\ae nge J D, Pisani P A G and Wipf A 2003 Algebraic Solution of the Supersymmetric Hydrogen Atom in d Dimension {\it Ann. Phys.} {\bf 303} 359-388.

\bibitem{Wipf2} Wipf A, Kirchberg A and L\ae nge D 2006 {\it Proc. Int. Symp. Quantum Theory and Symmetries (Varna)} (Sofia: Heron Press) p 887-897.


\end{thebibliography}
\end{document}